\documentclass[twocolumn,10pt,twoside]{IEEEtranTCOM}
\normalsize

\ifCLASSINFOpdf
\else
\fi

\usepackage{amsfonts}
\usepackage{amsmath}
\usepackage{graphicx}
\usepackage{epstopdf}
\usepackage{color}
\usepackage{multicol}
\usepackage{amssymb}
\usepackage{subcaption}
\usepackage{bm}
\usepackage{latexsym}
\usepackage{stfloats}
\usepackage{enumerate}

\usepackage{algorithm}
\usepackage{listings}
\usepackage{algpseudocode}
\usepackage{cases}

\begin{document}
%
\title{Hybrid Block Diagonalization for Massive Multiuser MIMO Systems}

%
%
%

\author{
Weiheng~Ni 
and~Xiaodai~Dong
\thanks{
W. Ni and X. Dong are with the Department of Electrical and Computer Engineering, University
of Victoria, Victoria, BC V8W 3P6, Canada (email: nweiheng@uvic.ca, xdong@ece.uvic.ca).
}
}

\maketitle
\newtheorem{theorem}{Theorem}[section]
\newtheorem{lemma}[theorem]{\textbf{Lemma}}
\newtheorem{proposition}[theorem]{Proposition}
\newtheorem{remark}[theorem]{\textbf{Remark}}
\newtheorem{corollary}[theorem]{Corollary}

\begin{abstract}
For a massive multiple-input multiple-output (MIMO) system, restricting the number of RF chains to far less than the number of antenna elements can significantly reduce the implementation cost compared to the full complexity RF chain configuration. In this paper, we consider the downlink communication of a massive multiuser MIMO (MU-MIMO) system and propose a low-complexity hybrid block diagonalization (Hy-BD) scheme to approach the capacity performance of the traditional BD processing method. We aim to harvest the large array gain through the phase-only RF precoding and combining and then digital BD processing is performed on the equivalent baseband channel. The proposed Hy-BD scheme is examined in both the large Rayleigh fading channels and millimeter wave (mmWave) channels. A performance analysis is further conducted for single-path channels and large number of transmit and receive antennas. Finally, simulation results demonstrate that our Hy-BD scheme, with a lower implementation and computational complexity,  achieves a capacity performance that is close to (sometimes even higher than) that of the traditional high-dimensional BD processing. 
\end{abstract}

\begin{IEEEkeywords}
Massive MIMO, large scale MU-MIMO, hybrid processing, block diagonalization, limited RF chains, mmWave
\end{IEEEkeywords}

%
\IEEEpeerreviewmaketitle

\section{Introduction}\label{sec:intro}
To realize the tremendous capacity target of the next generation mobile cellular systems, one promising option is scaling up to massive multiple-input multiple-output (MIMO) systems \cite{andrews2014what5g}-\cite{marzetta2010noncoop_mimo}. In the massive multiuser MIMO (MU-MIMO) systems, some simple linear pre/post-processing (transmit precoding/receive combining) schemes, such as zero-forcing (ZF) and linear minimum mean-square error (MMSE), are able to approach the optimal capacity performance achieved by the dirty paper coding (DPC) as the number of antennas goes to infinity \cite{erez2005dpc}. Moreover, the ZF processing that cancels the inter-user interference through channel inversion can be generalized as block diagonalization (BD) when the base stations (BSs) and mobile stations (MSs) are both equipped with multiple antennas \cite{spencer2004bd}. For the downlink spatial multiplexing in MU-MIMO systems, the BD method achieves sub-optimal capacity performance; however, it reduces the complexity of the transmitter and receiver structures by providing closed-form precoder and combiner solutions. From a different perspective, the problems of the downlink beamformer design for signal-to-interference-plus-noise ratio balancing and the downlink physical layer multicasting that aims at minimizing the transmit power in massive MIMO systems have been investigated in \cite{hanif2014beam} and \cite{hanif2014conic} respectively. Reference \cite{peng2010las} presents a low-complexity algorithm for detection in massive MIMO systems based on the likelihood ascent search (LAS) algorithm.

In large-scale MIMO systems, the large array gain is rendered by a massive number of antennas at the order of a hundred or more \cite{rusek2013scalemimo}. Conventional pre-processing is performed through modifying the amplitudes and phases of the complex transmit symbols at the baseband and then upconverted to the passband by going through radio frequency (RF) chains (including the digital-to-analog conversion, signal mixing and power amplifying), which requires that the number of the RF chains is in the range of hundreds, equal to the number of the antenna elements. Post-processing is similar involving a large number of analog receive RF chains and digital baseband operations. This leads to unacceptably high implementation cost and energy consumption. 

Recently, enabled by the cost-effective variable phase shifters, a limited number of RF chains have been applied in the MIMO systems \cite{love2003equalgain}-\cite{striling2014hybrid}. The analog RF processing provides the high-dimensional phase-only control while the digital baseband processing can be performed in a very low dimension, termed as hybrid processing.
Under the limited RF chains constraint, references \cite{love2003equalgain} and \cite{zhang2005antselect} investigate the hybrid processing schemes in the point-to-point (P2P) MIMO systems. A single-stream communication under the Rayleigh fading MIMO channels achieves the full diversity order through the equal gain transmission/combining (EGT/EGC) in \cite{love2003equalgain}, while the multiple-stream transmission under MIMO channels is proposed in \cite{zhang2005antselect}. In addition, \cite{liang2014hybrid} and \cite{liu2014rfonly} implement the hybrid processing to the downlink of the massive MU-MIMO systems with single-antenna users. In \cite{liang2014hybrid}, the near-optimal capacity performance, compared to the full-complexity systems, is achieved through the ZF baseband precoding combined with the EGT processing in the RF domain. Note that this technique also works for the millimeter wave (mmWave) channel. In \cite{liu2014rfonly}, the phase-only RF precoding are employed to maximize the minimum average data rate of users via a bi-convex approximation approach.

Furthermore, in mmWave communications systems, it is possible to build a large antenna array in a compact region and apply hybrid processing technique \cite{heath2012beamsteer}-\cite{alkh2014limited}. The ``dominant'' paths in P2P mmWave channels are captured through the hybrid processing in \cite{heath2012beamsteer} and \cite{heath2014sparse}, where the former considers the single-stream transmission while the latter enables the multiple-stream communication. \cite{heath2014sparse} presents a hybrid processing by decomposing the optimal precoding/combining matrix via orthogonal matching pursuit with the transmit/receive array response vectors as the basis vectors. Reference \cite{heath2012beamsteer} can be regarded as a special one-RF-chain case of reference \cite{heath2014sparse}.
On the other hand, in the mmWave MU-MIMO systems, \cite{sayeed2013beamspace} considers the single-antenna users and designs the analog RF precoding based on the transmit beam directions, while the digital processing (matched filter, zero-forcing or Wiener filter) performs on the baseband equivalent channels. With the multiple-antenna users, some baseband processing schemes such MMSE and BD are examined in \cite{striling2014hybrid}, which, however, neglects the design of the analog RF processing. In addition, a comprehensive limited feedback hybrid precoding scheme is proposed to configure hybrid precoders at the transmitter and analog combiners with a small training and feedback overhead, which is also effective for multiple-antenna users who have only one RF chain \cite{alkh2014limited}.

In this paper, we consider the downlink communication of a massive MU-MIMO system where the BS and all MSs have multiple antennas. With a limited number ($\geq 1$) of RF chains in BS and MSs, hybrid processing is applied as an alternative to the traditional high-cost full dimensional RF and baseband processing. 
We propose to utilize the RF precoding and combining to harvest the large array gain provided by the large number of antennas in the massive MU-MIMO channels, which shares the similar objective with the above references that study the hybrid processing in the MU-MIMO systems. However, the analog RF processing design for the MU-MIMO systems with multiple-antenna MSs accommodating multiple data streams per MS is not available in the literature and the novel BS RF precoder design is based on a newly defined ``aggregate intermediate channel''. More specifically, the RF combiners of all the MSs are obtained by selecting some of the discrete Fourier transform (DFT) bases, while the RF precoder of the BS is designed by extracting the phases of the conjugate transpose of the aggregate intermediate channel which incorporates the MS RF combiners and the original downlink channels. With the designed RF precoder and combiners, a low-dimensional BD processing can be performed at the baseband to cancel the inter-user interference, and the whole operation is named the hybrid BD (Hy-BD) scheme. The advantages of such a Hy-BD scheme can be summarized as follows:
\begin{itemize}
    \item[1)] Low implementation cost and low computation complexity;
    \item[2)] Applicability to both Rayleigh fading and mmWave massive MU-MIMO channels. Channel state information (CSI) is required but not the information of each individual propagation path;
    \item[3)] Reduction on the feedback overhead of the RF domain operations.
\end{itemize}

Simulation results demonstrate that the proposed Hy-BD scheme achieves a capacity performance that is quite close to, sometimes even higher than, that of the full-complexity BD scheme in \cite{spencer2004bd} with a lower implementation and computational cost. The Hy-BD scheme is also examined in the mmWave MU-MIMO communication channels and compared to the spatially sparse precoding/combining method \cite{heath2014sparse} initially proposed for SU-MIMO but extended to MU-MIMO in this paper.

\section{System Model}\label{sec:sys_model}
\subsection{System Model}
\begin{figure}[htbp]
    \centering
    \includegraphics[width=0.45\textwidth]{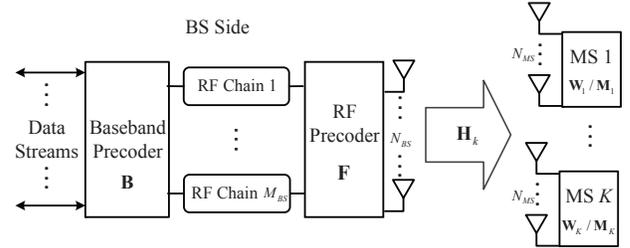}
    \caption{System diagram of a massive MU-MIMO system with hyrbid processing structure.}
    \label{fig:system_model}
\end{figure}

We consider the downlink communication of a massive multiuser MIMO system shown in Fig. \ref{fig:system_model}, where a base station with $N_{BS}$ antennas and $M_{BS}$ RF chains is assumed to schedule $K$ mobile stations. Each MS is equipped with $N_{MS}$ antennas and $M_{MS}$ RF chains to support $N_S$ data streams, which means total $KN_S$ data streams are handled by the BS. To guarantee the effectiveness of the communication carried by the limited number of RF chains, the number of the transmitted steams is constrained by $KN_S \leq M_{BS} \leq N_{BS}$ for the BS and $N_S \leq M_{MS} \leq N_{MS}$ for each MS.

At the BS, the transmitted symbols are assumed to be processed by a baseband precoder $\mathbf{B}$ of dimension $M_{BS} \times KN_S$ and then by an RF precoder $\mathbf{F}$ of dimension $N_{BS} \times M_{BS}$. Notably, the baseband precoder $\mathbf{B}$ enables both amplitude and phase modification, while only phase changes (phase-only control) can be realized by $\mathbf{F}$ since it is implemented by using analog phase shifters. Each entry of $\mathbf{F}$ is normalized to satisfy $|\mathbf{F}^{(i,j)}| = \frac{1}{\sqrt{N_{BS}}}$, where $|\mathbf{F}^{(i,j)}|$ denotes the amplitude of the $(i, j)$-th element of $\mathbf{F}$. Furthermore, to meet the total transmit power constraint, $\mathbf{B}$ is normalized to satisfy $||\mathbf{F}\mathbf{B}||_F^2 = KN_S$, where $||\cdot||_F$ the Frobenius norm.

We assume a narrowband flat fading channel model and obtain the received signal of the $k$-th MS
\begin{equation}
    \mathbf{y}_k = \mathbf{H}_k \mathbf{F}\mathbf{B} \mathbf{s} + \mathbf{n}_k,~k=1,2,\cdots,K,
\end{equation}
where $\mathbf{s} \in \mathbb{C}^{KN_S \times 1}$ is the signal vector for a total of $K$ MSs, each of which processes a $N_S \times 1$ signal vector $\mathbf{s}_k$. Namely, $\mathbf{s} = [\mathbf{s}_1^T, \mathbf{s}_2^T, \cdots, \mathbf{s}_K^T]^T$, where $(\cdot)^T$ denotes transpose. And the signal vector satisfies $\mathbb{E}[\mathbf{ss}^H] = \frac{P}{KN_{S}}\mathbf{I}_{KN_{S}}$, where $(\cdot)^H$ denotes conjugate transpose, $\mathbb{E}[\cdot]$ denotes expectation, $P$ is the average transmit power and $\mathbf{I}_{KN_S}$ is the $KN_S \times KN_S$ identity matrix. $\mathbf{H}_k \in \mathbb{C}^{N_{MS} \times N_{BS}}$ is the channel matrix for the $k$-th MS, and $\mathbf{n}_k$ is the $N_{MS} \times 1$ vector of i.i.d. $\mathcal{CN}(0, \sigma^2)$ additive complex Gaussian noise. And the processed received signal at the $k$-th MS after combining is given by
\begin{equation}
    \tilde{\mathbf{y}}_k = \mathbf{M}_k^H\mathbf{W}_k^H \mathbf{H}_k \mathbf{F}\mathbf{B} \mathbf{s} + \mathbf{M}_k^H\mathbf{W}_k^H\mathbf{n}_k,~k=1,2,\cdots,K,
\end{equation}
where $\mathbf{W}_k$ is the $N_{MS} \times M_{MS}$ RF combining matrix and $\mathbf{M}_k$ is the $M_{MS} \times N_S$ baseband combining matrix for the $k$-th MS. Since $\mathbf{W}_k$ is also implemented by the analog phase shifters, all elements of $\mathbf{W}_k$ should have the constant amplitude such that $|\mathbf{W}_k^{(i,j)}| = \frac{1}{\sqrt{N_{MS}}}$. We define an equivalent baseband channel for each MS as
\begin{equation}\label{eq:ch_eq_one_user}
    \tilde{\mathbf{H}}_k = \mathbf{W}_k^H\mathbf{H}_k \mathbf{F},~k=1,2,\cdots,K,
\end{equation}
and the entire equivalent multiuser baseband channel can be denoted as
\begin{equation}\label{eq:ch_eq_all_user}
    \mathbf{H}_{eq} = \left[                 
    \begin{array}{c}   
        \tilde{\mathbf{H}}_1 \\ 
        \tilde{\mathbf{H}}_2 \\
        \vdots \\
        \tilde{\mathbf{H}}_K
    \end{array}
    \right]   
    = \left[
    \begin{array}{cccc}   
        \mathbf{W}_1^H & \mathbf{0} & \cdots & \mathbf{0} \\ 
        \mathbf{0} & \mathbf{W}_2^H & \cdots & \mathbf{0} \\
        \vdots & \vdots & \ddots & \vdots \\
        \mathbf{0} & \mathbf{0} & \cdots & \mathbf{W}_K^H
    \end{array}
    \right]
    \left[                 
    \begin{array}{c}   
        \mathbf{H}_1 \\ 
        \mathbf{H}_2 \\
        \vdots \\
        \mathbf{H}_K
    \end{array}
    \right] \mathbf{F}.
\end{equation}
Then the processed received signal at the $k$-th MS can also be represented as
\begin{equation}\label{eq:recv_sig_processed}
\begin{aligned}
    \tilde{\mathbf{y}}_k &= \mathbf{M}_k^H \tilde{\mathbf{H}}_k \mathbf{B}_k \mathbf{s}_k 
    + \underbrace{\sum_{i=1, i\neq k}^{K}\mathbf{M}_k^H \tilde{\mathbf{H}}_k \mathbf{B}_i \mathbf{s}_i}_{\mathrm{interference}} \\
    &+ \underbrace{\mathbf{M}_k^H\mathbf{W}_k^H\mathbf{n}_k}_{\mathrm{noise}},~k=1,2,\cdots,K,
\end{aligned}
\end{equation}
where $\mathbf{B}_k$ is the $((k-1)N_S + 1)$-th to the $(kN_S)$-th columns of $\mathbf{B}$, corresponding to the baseband precoding for $\mathbf{s}_k$. When the Gaussian symbols are used by the BS, the sum spectral efficiency achieved will be
\begin{equation}
    R = \sum_{k=1}^K \log_2 \left( \left| \mathbf{I}_{N_S} + \frac{P}{KN_S}\mathbf{R}_{i}^{-1} \mathbf{M}_k^H \tilde{\mathbf{H}}_k \mathbf{B}_k \mathbf{B}_k^H \tilde{\mathbf{H}}_k^H  \mathbf{M}_k\right| \right),
\end{equation}
where $\mathbf{R}_{i} = \frac{P}{KN_S}\sum_{i=1, i\neq k}^{K}\mathbf{M}_k^H \tilde{\mathbf{H}}_k \mathbf{B}_i \mathbf{B}_i^H \tilde{\mathbf{H}}_k^H \mathbf{M}_k + \sigma^2\mathbf{M}_k^H\mathbf{W}_k^H\mathbf{W}_k\mathbf{M}_k$ is the covariance matrix of both interference and noise. 

Generally, joint optimization on the RF and baseband precoders and combiners should be an essential method to design the processing scheme that achieves optimal sum spectral efficiency $R$. However, as stated in \cite{heath2014sparse}, finding global optima for similar constrained joint optimization problems (maxmizing $R$ while constant-amplitude contraints imposed to the RF analog precoder and combiners) is often found to be intractable. Even in the traditional MU-MIMO systems without hybrid processing structure, it also needs enormous efforts to find a local optimum of sum rate by alternating optimization \cite{weighted2008christensen}. For some recently designed hybrid processing schemes \cite{liang2014hybrid}\cite{striling2014hybrid}-\cite{alkh2014limited} in the literature, separated RF and baseband processing designs are investigated to obtain satisfying performance without involving a myriad of iterative procedures. Therefore, we choose to separate the RF and baseband domain designs in this paper.

\subsection{Channel Model}
In this paper, the general channel matrix is set as $\mathbf{H} = [\mathbf{H}_1^T, \mathbf{H}_2^T, \cdots, \mathbf{H}_K^T]^T = [\sqrt{\beta_1}\mathbf{\dot{H}}_1^T, \sqrt{\beta_2}\mathbf{\dot{H}}_2^T, \cdots, \sqrt{\beta_K}\mathbf{\dot{H}}_K^T]^T$, where $\sqrt{\beta_k}$ and $\mathbf{\dot{H}}_k$ indicate the large scale path fading and normalized channel matrix respectively for the $k$-th MS, satisfying that $\mathbb{E}[||{\mathbf{\dot{H}}_k}||_F^2] = N_{BS}N_{MS}$. With the knowledge of the general channel matrix, we aim to seek the BS hybrid precoders $(\mathbf{F}$, $\mathbf{B})$ and the hybrid combiners $(\mathbf{W}_k$, $\mathbf{M}_k)$'s for all $K$ MSs through the Hy-BD scheme, which achieves a sub-optimal spectral efficiency for massive MU-MIMO systems by perfectly canceling the inter-user interference. Two kinds of channel models are considered in this paper:
\begin{itemize}
    \item[1)] large i.i.d. Rayleigh fading channel $\mathbf{H}_{rl}$;
    \item[2)] limited scattering mmWave channel $\mathbf{H}_{mmw}$.
\end{itemize}

In the large Rayleigh fading channel, which is commonly considered in massive MU-MIMO systems, all entries of the normalized channel matrix $\mathbf{\dot{H}}_k$ for the $k$-th MS follow i.i.d. $\mathcal{CN}(0, 1)$. On the other hand, a large antenna array is often implemented in mmWave communications to combat the high free-space pathloss \cite{heath2012beamsteer}-\cite{sayeed2013beamspace}. We adopt the clustered mmWave channel model to characterize the limited scattering feature of the mmWave channel. The normalized mmWave downlink channel for the $k$-th MS $\mathbf{\dot{H}}_k$ is assumed to be the sum of all propagation paths that are scattered in $N_c$ clusters and each cluster contributes $N_p$ paths, which can be expressed as
\begin{equation}\label{eq:cl_ch_model}
    \mathbf{\dot{H}}_k = \sqrt{\frac{N_{BS}N_{MS}}{N_cN_p}}\sum_{i = 1}^{N_c}\sum_{l=1}^{N_p}\alpha_{il}^k \mathbf{a}_{MS}^k(\theta_{il}^k) \mathbf{a}_{BS}^k(\phi_{il}^k)^H,
\end{equation}
where $\alpha_{il}^k$ is the complex gain of the $i$-th path in the $l$-th cluster, which follows $\mathcal{CN}(0, 1)$.
To reflect the sparsity of the mmWave channel, both of $N_c$ and $N_p$ should not be too large.
 For the $(i, l)$-th path, $\theta_{il}^k$ and $\phi_{il}^k$ are the azimuth angles of arrival/departure (AoA/AoD), while $\mathbf{a}_{MS}^k(\theta_{il}^k)$ and $\mathbf{a}_{BS}^k(\phi_{il}^k)$ are the receive and transmit array response vectors at the azimuth angles of $\theta_{il}^k$ and $\phi_{il}^k$ respectively, and the elevation dimension is ignored. 
Within the cluster $i$, $\theta_{il}^k$ and $\phi_{il}^k$ have the uniformly-distributed mean values of $\theta_i^k$ and $\phi_i^k$ respectively, while the lower and upper bounds of the uniform distribution for $\theta_i^k$ and $\phi_i^k$ can be defined as $[\theta_{\min}^k, \theta_{\max}^k]$ and $[\phi_{\min}^k, \phi_{\max}^k]$. The angle spreads (standard deviations) of $\theta_{il}^k$ and $\phi_{il}^k$ among all clusters are assumed to be constant, denoted as $\sigma_\theta^k$ and $\sigma_\phi^k$. Finally, the truncated Laplacian distribution is employed to generate all the AoDs/AoAs for this mmWave propagation channel matrix, base on the above parameters. 

The uniform linear array (ULA) is employed by the BS and MSs in our study, while the Hy-BD scheme in Section-\ref{sec:hy_bd_scheme} can directly be applied to arbitrary antenna arrays. For an N-element ULA, the array response vector can be given by
\begin{equation}\label{eq:ula}
    \mathbf{a}_{ULA}(\theta) = \frac{1}{\sqrt{N}}\left[ 1, e^{j\frac{2\pi}{\lambda}d\sin(\theta)}, \cdots, e^{j(N-1)\frac{2\pi}{\lambda}d\sin(\theta)} \right]^T,
\end{equation}
where $\lambda$ is the wavelength of the carrier, and $d$ is the distance between neighboring antenna elements. The array response vectors of the BS and MSs can be written in the form of (\ref{eq:ula}). Furthermore, other non-ULA antenna geometries, such as uniform planar array (UPA), are also examined in the simulations.

\section{Hybrid Block Diagonalization}\label{sec:hy_bd_scheme}
In the MU-MIMO systems, the generalized zero-forcing method (i.e., the traditional BD scheme) is infeasible to be practically implemented due to the high cost brought by the large number of RF chains as many as the antennas. By reducing the number of RF chains $M_{BS} (M_{MS})$ to far less than the antenna elements $N_{BS} (N_{MS})$ at both the BS and MSs, we propose to utilize the RF precoding matrix $\mathbf{F}$ at the BS and the RF combining matrix $\mathbf{W}_k$ at each MS to harvest the large array gain provided by the large number of antennas in the massive MU-MIMO channel. With the found $\mathbf{F}$ and all $\mathbf{W}_k$'s, the entire multiuser equivalent baseband channel $\mathbf{H}_{eq}$ can be determined based on (\ref{eq:ch_eq_all_user}), which consists of all the equivalent channels for the MSs, namely $\tilde{\mathbf{H}}_k,~ k=1,2,\cdots,K$. Finally, a low-dimensional BD processing, involving the design of $\mathbf{B}$ and all $\mathbf{M}_k$'s, can be performed at the baseband.

\subsection{Array Gain Harvesting}\label{sec:array_gain}
Owing to the large number of antennas in the massive MU-MIMO systems, the channel gains of the equivalent channel $\mathbf{H}_{eq}$ can be scaled up through the appropriate phase-only control at the RF domain, which is called the large array gain. To be noted, each element in $\mathbf{H}_{eq}$ represents the equivalent channel gain from one RF chain at the BS to one RF chain at one MS. To achieve the high capacity with such a hybrid processing structure, the equivalent channel matrix $\mathbf{H}_{eq}$ are desired to have the following properties:
\begin{itemize}
    \item[1)] Rank sufficiency: $\mathbf{H}_{eq}$ should be well-conditioned to support the multi-stream transmission, which means the rank of $\mathbf{H}_{eq}$ should be at least $KN_S$;
    \item[2)] Large array gain: $\mathbf{H}_{eq}$ should sufficiently harvest the array gain so that it can provide as large gain for each stream transmission as possible. We propose to pursue the large array gain by enlarging the sum of the squares of the diagonal entries in $\mathbf{H}_{eq}$.
\end{itemize}
By definition, $\mathbf{H}_{eq}$ consists of the equivalent channels of all the MSs, namely $\tilde{\mathbf{H}}_k = \mathbf{W}_k^H\mathbf{H}_k \mathbf{F},~k=1,2,\cdots,K$. We design the RF domain processing matrices $\mathbf{W}_k$'s and $\mathbf{F}$ and construct the equivalent channel $\mathbf{H}_{eq}$ by approximately satisfying the above two requirements, which will lead to a suboptimal performance under the hybrid precoding structure, but with significantly low complexity. 

Assume that all the RF combiners $\mathbf{W}_k$'s are given (the actual design of $\mathbf{W}_k$'s will be presented shortly). Define an aggregate intermediate channel given by
\begin{equation}\label{eq:ch_inter}
    \mathbf{H}_{int} = \left[                 
    \begin{array}{c}   
        \mathbf{W}_1^H\mathbf{H}_1 \\ 
        \vdots \\
        \mathbf{W}_K^H\mathbf{H}_K
    \end{array}
    \right]_{KM_{MS} \times N_{BS}},
\end{equation}

and then the baseband equivalent channel is $\mathbf{H}_{eq} = \mathbf{H}_{int}\mathbf{F}$. Due to the phase shifting ability of the RF precoder and the knowledge of the channel matrix entries, we perform the phase-only RF precoding based on an equal gain transmission (EGT) method proposed in \cite{liang2014hybrid} to harvest the large array gain, by setting
\begin{equation}
    \mathbf{F}^{(i,j)} = \frac{1}{\sqrt{N_{BS}}}e^{j\psi_{i,j}},
\end{equation}
where $\psi_{i,j}$ is the phase of the $(i,j)$-th element of the conjugate transpose of $\mathbf{H}_{int}$. This EGT precoding method requires $M_{BS} = KM_{MS}$ RF chains at the BS, which means $\mathbf{F}$ is an $N_{BS} \times KM_{MS}$ matrix and $\mathbf{H}_{eq}$ should be a square matrix. The entries along the diagonal of the baseband equivalent channel $\mathbf{H}_{eq}$ denote the equivalent channel gains in terms of the RF chains, while the remaining entries indicate the inter-chain interference. We focus on the large array gain design through the RF precoding/combining and leave  the interference canceling to the baseband processing in the Hy-BD scheme. 

Now let us return to the design of the RF combiners $\mathbf{W}_k$'s. Denote the $m$-th column of $\mathbf{W}_k$ as $\mathbf{w}_k^{(m)}$. As the result of the EGT precoding method, the $((k-1)M_{MS} + m)$-th diagonal entry of $\mathbf{H}_{eq}$ is then given by $||(\mathbf{w}_k^{(m)})^H \mathbf{H}_k||_1$, where $||\cdot||_1$ denotes the 1-norm of a vector, corresponding to the $m$-th RF chain of the $k$-th MS. Note that the entries in $\mathbf{H}_{eq}$ indicate the RF-chain to RF-chain channel gains and those off-diagonal entries indicate the inter-RF-chain, and even inter-user, interference. We aim to maximize the sum of the squares of diagonal entries of the baseband equivalent channel $\mathbf{H}_{eq}$, given by $\sum_{k = 1}^{K}\sum_{m=1}^{M_{MS}}||(\mathbf{w}_k^{(m)})^H\mathbf{H}_k||_1^2$, to pursue the large array gain. Due the independence of $\mathbf{W}_k$'s for all the MSs, maximizing $\sum_{k = 1}^{K}\sum_{m=1}^{M_{MS}}||(\mathbf{w}_k^{(m)})^H\mathbf{H}_k||_1^2$ is equivalent to maximizing $\sum_{m=1}^{M_{MS}}||(\mathbf{w}_k^{(m)})^H\mathbf{H}_k||_1^2$ for all $k = 1, \cdots, K$ respectively. Hence, the design of the RF combiners can be obtained by solving
\begin{equation}\label{eq:wk_problem}
\begin{aligned}
    \max_{\mathbf{W}_k}~&\sum_{m=1}^{M_{MS}}||(\mathbf{w}_k^{(m)})^H\mathbf{H}_k||_1^2 \\
    s.t.~&|\mathbf{W}_k^{(i,j)}| = \frac{1}{\sqrt{N_{MS}}},~\forall i,j.
\end{aligned}
\end{equation}
Herein, we need to clarify that no inter-user interference is designed to be suppressed by solving the simplified maximization problem in (\ref{eq:wk_problem}), which, as a heuristic method, does not guarantee the optimality of the sum-rate maximization, but lend tractability to approaching a sub-optimal solution. In this paper, instead of solving the non-convex problem (\ref{eq:wk_problem}) directly, we modify the constraints to choose from a set of DFT basis, as explained in details next. Note that $||(\mathbf{w}_k^{(m)})^H\mathbf{H}_k||_1^2 = (\sum_{n=1}^{N_{BS}}|(\mathbf{w}_k^{(m)})^H\mathbf{h}_k^{(n)}|)^2$, where $\mathbf{h}_k^{(n)}$ denotes the $n$-th column of $\mathbf{H}_k$. Moreover, the geometric MIMO channel models, including the Rayleigh fading\footnote{In the Rayleigh fading channel, all AoDs/AoAs of the paths (non-LOS) are uniformly distributed among $[0, 2\pi)$ and the number of paths approaches to infinity.} and mmWave channels, can be represented in the form of (\ref{eq:cl_ch_model}), which means $\mathbf{h}_k^{(n)}$ is the linear combination of all the array response vectors of the AoAs. This fact implies that each addition term in $||(\mathbf{w}_k^{(m)})^H\mathbf{H}_k||_1$, $|(\mathbf{w}_k^{(m)})^H\mathbf{h}_k^{(n)}|$ is the absolute value of the summing weighted projections of those array response vectors $\mathbf{a}_{MS}^k(\theta_{il}^k)$ for all AoAs onto $\mathbf{w}_k^{(m)}$. From this perspective, we first propose to set $\mathbf{w}_k^{(m)}$ in the form of array response vector (\ref{eq:ula}) to extract the gain from these projections, namely,
\begin{equation}\label{eq:wk_dtf_form}
    \mathbf{d}(\omega) = \frac{1}{\sqrt{N_{MS}}}\left[ 1, e^{j\omega}, e^{j2\omega}, \cdots, e^{j(N_{MS}-1)\omega} \right]^T,
\end{equation}
where $\omega = \frac{2\pi}{\lambda}d\sin\theta$ denotes the corresponding spatial frequency \cite{dinesh2012csadaption}.

Furthermore, to meet the rank sufficiency requirement of $\mathbf{H}_{eq}$, it is desirable that the rank of $\mathbf{H}_k$ is not reduced after it being multiplied by $\mathbf{W}_k$. For this purpose, we require the columns of $\mathbf{W}_k$ to be pairwise orthogonal so that the rank of $\mathbf{W}_k^H\mathbf{H}_k$ is lower bounded by $M_{MS} > N_S$ (the rank of the high-dimensional $\mathbf{H}_k$ is assumed to be larger than $M_{MS}$), which means the equivalent channel $\mathbf{H}_{eq}$ is potentially capable of supporting the transmission of $KM_{MS} > KN_S$ streams. Considering the form of $\mathbf{w}_{k}^{(m)}$, we discretize the $\omega$ into $N_{MS}$ levels over $[0, 2\pi)$ and construct $N_{MS}$ bases, given by $\mathbf{D} = \{\mathbf{d}(0), \mathbf{d}(\frac{2\pi}{N_{MS}}), \cdots, \mathbf{d}(\frac{2\pi(N_{MS}-1)}{N_{MS}})\}$ as the candidates from which the $\mathbf{w}_k^{(m)}$ is choosen. As we can see, these bases in $\mathbf{D}$ exactly form an $N_{MS}$-dimensional DFT basis set, which simultaneously conforms to the rank sufficiency and large arrary gain requirements of $\mathbf{H}_{eq}$. Therefore, we finally design the RF combiners by solving
\begin{equation}\label{eq:wk_pro_dft_cons}
\begin{aligned}
    \max_{\mathbf{W}_k}~& \sum_{m=1}^{M_{MS}}||({\mathbf{w}_k^{(m)}})^H\mathbf{H}_k||_1^2 \\
    s.t.~&\mathbf{w}_k^{(m)} \in \mathbf{D},~m = 1,\cdots, M_{MS}.
\end{aligned}
\end{equation}
To solve the problem (\ref{eq:wk_pro_dft_cons}), we just need to sort all $N_{MS}$ $||\mathbf{d}(\omega)^H\mathbf{H}_k||_1$'s in the descending order and then choose the first $M_{MS}$ $\mathbf{d}(\omega)$'s as the columns of $\mathbf{W}_k$. Note that each MS only needs to solve problem (\ref{eq:wk_pro_dft_cons}) with the corresponding index $k$ for once to obtain its RF combiner. In addition, the number of antennas $N_{MS}$ of an MS usually is much smaller than $N_{BS}$ due to the actual device size and computational capacity, which makes the exhaustive search on the DFT bases acceptable.

\begin{remark}
Based on the selection of the DFT bases, the MSs can avoid a huge amount of computation overhead for obtaining all the phase shift elements. In addition, only $N_S$ phase shift elements per MS is needed to be fed back to the BS, so that to the BS is able to re-construct all the $\mathbf{W}_k$'s and calculate the aggregate intermediate channel $\mathbf{H}_{int}$ for further processing. 
\end{remark}

\subsection{Baseband Block Diagonalization}
In this section, based on the obtained baseband equivalent channel $\mathbf{H}_{eq}$, given the found RF processing matrices $\mathbf{W}_k$ and $\mathbf{F}$, we perform the low-dimensional BD processing with the baseband precoder $\mathbf{B}$ and combiners $\mathbf{M}_k$'s to cancel the inter-user interference, which forces the interference terms $\tilde{\mathbf{H}}_k \mathbf{B}_i = \mathbf{0}$ for $i \neq k$ in (\ref{eq:recv_sig_processed}). The spectral efficiency of the MU-MIMO system can be further simplified to
\begin{equation}
\begin{aligned}
    R = \sum_{k=1}^K \log_2 \left( \left| \mathbf{I}_{N_S} + \frac{P}{\sigma^2KN_S}(\mathbf{M}_k^H\mathbf{W}_k^H\mathbf{W}_k\mathbf{M}_k)^{-1} \right.\right.\\
    \left.\left. \mathbf{M}_k^H \tilde{\mathbf{H}}_k \mathbf{B}_k \mathbf{B}_k^H \tilde{\mathbf{H}}_k^H  \mathbf{M}_k\right| \right).
\end{aligned}
\end{equation}

To obtain the baseband precoder $\mathbf{B} = [\mathbf{B}_1, \mathbf{B}_2, \cdots, \mathbf{B}_K]$, where $\mathbf{B}_k$ incorporates the precoding vectors for the data streams of the $k$-th MS, we first define $\overline{\mathbf{H}}_k$ as
\begin{equation}
    \overline{\mathbf{H}}_k = [\tilde{\mathbf{H}}_1^T, \cdots, \tilde{\mathbf{H}}_{k-1}^T, \tilde{\mathbf{H}}_{k+1}^T, \cdots, \tilde{\mathbf{H}}_K^T]^T.
\end{equation}
The $\mathbf{B}_k$ is supposed to lie in the null space of $\overline{\mathbf{H}}_k$. Denote the rank of $\overline{\mathbf{H}}_k$ as $r_k \leq (K-1)M_{MS}$. Then the singular value decomposition (SVD) of $\overline{\mathbf{H}}_k$ is given by
\begin{equation}
    \overline{\mathbf{H}}_k = \overline{\mathbf{U}}_k \overline{\mathbf{\Sigma}}_k \left[\overline{\mathbf{V}}_k^{((K-1)M_{MS})},~\overline{\mathbf{V}}_k^{(M_{MS})}\right]^H,
\end{equation}
where $\overline{\mathbf{V}}_k^{((K-1)M_{MS})}$ consists of the first $(K-1)M_{MS}$ right singular vectors of $\overline{\mathbf{H}}_k$, and $\overline{\mathbf{V}}_k^{(M_{MS})}$ holds the rest $M_{MS}$ ones which are exactly the orthogonal bases of the null space of $\overline{\mathbf{H}}_k$. Then we know 
\begin{equation}
    \tilde{\mathbf{H}}_i\overline{\mathbf{V}}_k^{(M_{MS})} = \left\{
    \begin{array}{rl}
        \mathbf{0},~i \neq k\\
        \tilde{\mathbf{H}}_k\overline{\mathbf{V}}_k^{(M_{MS})},~i=k
    \end{array}
    \right.
\end{equation}

Given the above results, block diagonalization of the baseband equivalent channel matrix to remove inter-user interference is written as
\begin{equation}
\begin{aligned}
    \mathbf{H}_{BD} &= \mathbf{H}_{eq}\left[\overline{\mathbf{V}}_1^{(M_{MS})},\cdots,\overline{\mathbf{V}}_K^{(M_{MS})}\right] \\
    &= \left[
    \begin{array}{ccc}   
        \tilde{\mathbf{H}}_1\overline{\mathbf{V}}_1^{(M_{MS})} & \cdots & \mathbf{0} \\ 
        \vdots & \ddots & \vdots \\
        \mathbf{0} & \cdots & \tilde{\mathbf{H}}_K\overline{\mathbf{V}}_K^{(M_{MS})}
    \end{array}
    \right].
\end{aligned}
\end{equation}
Until now, all the MSs can perform interuser-interference-free multi-stream transmission through their own sub-channels (the non-zero block in $\mathbf{H}_{BD}$). Further precoding/combining will be performed to achieve each MS's optimal spectral efficiency based on SVD, given by
\begin{equation}
    \tilde{\mathbf{H}}_k\overline{\mathbf{V}}_k^{(M_{MS})}
    = \mathbf{U}_k \mathbf{\Sigma}_k \mathbf{V}_k^H.
\end{equation}
With the above rank sufficiency requirement, $\tilde{\mathbf{H}}_k\overline{\mathbf{V}}_k^{(M_{MS})}$ is a $M_{MS}$-by-$M_{MS}$ full-rank sub-channel matrix which enables $M_{MS} \geq N_S$ data streams transmission for the $k$-th MS. Therefore, the optimal precoder and combiner on the $k$-th effective sub-channel $\tilde{\mathbf{H}}_k\overline{\mathbf{V}}_k^{(M_{MS})}$ should be $\mathbf{V}_k^{(N_S)}$ and $\mathbf{U}_k^{(N_S)}$, where $\mathbf{V}_k^{(N_S)}$ and $\mathbf{U}_k^{(N_S)}$ are the first $N_S$ columns of the $\mathbf{V}_k$ and $\mathbf{U}_k$ respectively. Finally, the overall baseband precoder is given by
\begin{equation}
\begin{aligned}
    \mathbf{B} &= \left[\overline{\mathbf{V}}_1^{(M_{MS})},\cdots,\overline{\mathbf{V}}_K^{(M_{MS})}\right]
    \left[
    \begin{array}{ccc}   
        \mathbf{V}_1^{(N_S)} & \cdots & \mathbf{0} \\ 
        \vdots & \ddots & \vdots \\
        \mathbf{0} & \cdots & \mathbf{V}_K^{(N_S)}
    \end{array}
    \right] \\
    &= \left[\overline{\mathbf{V}}_1^{(M_{MS})}\mathbf{V}_1^{(N_S)},\cdots,\overline{\mathbf{V}}_K^{(M_{MS})}\mathbf{V}_K^{(N_S)}\right]_{KM_{MS} \times KN_S}.
\end{aligned}  
\end{equation}
And the baseband combiner for the $k$-th MS is given by $\mathbf{M}_k = \mathbf{U}_k^{(N_S)},~k=1,2,\cdots,K$. 

The spectral efficiency achieved by the Hy-BD scheme finally becomes
\begin{equation}\label{eq:pure_sum_rate}
\begin{aligned}
    R &= \sum_{k=1}^K \log_2 \left( \left| \mathbf{I}_{N_S} + \frac{P\mathbf{\Lambda}_k(\mathbf{M}_k^H\mathbf{W}_k^H\mathbf{W}_k\mathbf{M}_k)^{-1} (\mathbf{\Sigma}_k^{(N_S)})^2}{\sigma^2KN_S}\right| \right) \\
    &\stackrel{(\mathrm{i})}{=} \sum_{k=1}^K \log_2 \left( \left| \mathbf{I}_{N_S} + \frac{P\mathbf{\Lambda}_k (\mathbf{\Sigma}_k^{(N_S)})^2}{\sigma^2KN_S}\right| \right),
\end{aligned}
\end{equation}
where $\mathbf{\Lambda}=\mathrm{diag}\{\mathbf{\Lambda}_1, \mathbf{\Lambda}_2, \cdots, \mathbf{\Lambda}_K\}$ is a $KN_S \times KN_S$ diagonal matrix that performs water-filling power allocation, and $\mathbf{\Sigma}_k^{(N_S)}$ represents the first $N_S \times N_S$ block partition of $\mathbf{\Sigma}_k$. As we choose DFT bases (or any other orthogonal bases) to construct $\mathbf{W}_k$'s, the simplification step (i) of (\ref{eq:pure_sum_rate}) holds due to $\mathbf{M}_k^H\mathbf{W}_k^H\mathbf{W}_k\mathbf{M}_k = (\mathbf{U}_k^{(N_S)})^H\mathbf{U}_k^{(N_S)} = \mathbf{I}_{N_S}$.

On the other hand, since the RF and baseband processing do not require any information of the propagation paths of the channels $\mathbf{H}_k$'s, namely, each term in the summation of (\ref{eq:cl_ch_model}), the Hy-BD scheme can be performed on any kinds of massive MU-MIMO channels as long as the channel matrices are provided. 

\subsection{Proportional Water-Filling Power Allocation for Weighted Sum-Rate Maximization}
After employing the Hy-BD processing scheme, the optimal power allocation for transmitted data streams can be achieved by water-filling due to the sum-rate (sum-log) maximization form in (\ref{eq:pure_sum_rate}). However, considering the real-world scenarios, where fairness among users would often be considered, the pure sum-rate maximization in (\ref{eq:pure_sum_rate}) is not enough to guarantee the performance for some high-priority MSs or MSs located farther away from the BS. Therefore, the weighted sum-rate maximization is a more suitable objective when allocating transmission power to achieve proportional fairness. That is,
\begin{equation}\label{eq:weighted_k_sum_rate}
\begin{aligned}
	\max_{\mathbf{\Lambda}}~&R = \sum_{k=1}^K w_k \log_2 \left( \left| \mathbf{I}_{N_S} + \frac{P\mathbf{\Lambda}_k (\mathbf{\Sigma}_k^{(N_S)})^2}{\sigma^2KN_S}\right| \right) \\
	s.t.~&\mathrm{trace}\{\mathbf{\Lambda}\} = KN_S,\\
	&\mathbf{\Lambda}^{(n, n)} \ge 0,~\mathrm{for}~n = 1, \cdots, KN_S,
\end{aligned}
\end{equation}
where $w_k$ is the positive weight for the achievable rate of the $k$-th MS. Slightly abusing the notation, we write the $n$-th diagonal element of $\mathbf{\Lambda}$ and $\frac{P}{\sigma^2 KN_S}\mathrm{diag}\{(\mathbf{\Sigma}_1^{(N_S)})^2), (\mathbf{\Sigma}_2^{(N_S)})^2, \cdots, (\mathbf{\Sigma}_K^{(N_S)})^2\}$ as $\lambda_n$ and $\gamma_n$ respectively. Then (\ref{eq:weighted_k_sum_rate}) can be rewritten as a convex optimization problem
\begin{equation}\label{eq:weighted_sum_log_rate}
\begin{aligned}
	\min_{\{\lambda_n\}}~& -\sum_{n=1}^{KN_S} \tilde{w}_n \ln \left( 1 + \gamma_n \lambda_n \right) \\
	s.t.~&\sum_{n=1}^{KN_S}\lambda_n = KN_S, \\
	&\lambda_n \ge 0,~\mathrm{for}~n = 1, \cdots, KN_S,
\end{aligned}
\end{equation}
where $\tilde{w}_{(k-1)N_s + i} = w_k, k = 1, \cdots, K$ and $i = 1, \cdots, N_s.$ Similar to Example 5.2 in \cite{boyd2004opt}, we introduce Lagrange multipliers $\{m_1, \cdots, m_{KN_S}\} \in \mathbf{R}^{KN_S}$ for the inequality constraints $\lambda_n \ge 0$ and a multiplier $v \in \mathbf{R}$ for the equality constraints $\sum_n^{KN_S}\lambda_n = KN_S$, and the KKT conditions are
\begin{equation}\label{eq:kkt_conditions}
\begin{aligned}
	&\sum_{n=1}^{KN_S}\lambda_n = KN_S,~\lambda_n \ge 0,~m_n \ge 0,~m_n\lambda_n=0,\\
	&-\frac{\tilde{w}_n \gamma_n}{1+\gamma_n\lambda_n} - m_n + v = 0,~\mathrm{for}~n = 1, \cdots, KN_S.
\end{aligned}
\end{equation}
Then we can directly obtain that $\lambda_n m_n = \lambda_n\left(v-\frac{\tilde{w}_n \gamma_n}{1+\gamma_n\lambda_n}\right) = \frac{\lambda_n}{\tilde{w}_n}\left(v-\frac{1}{\frac{1}{\tilde{w}_n\gamma_n}+\frac{\lambda_n}{\tilde{w}_n}}\right)\textbf{•} = 0$, which results in
\begin{equation}\label{eq:proportional_wf}
    \lambda_n = \tilde{w}_n\max\{\frac{1}{v} - \frac{1}{\tilde{w}_n\gamma_n}, 0\} = \max\{\frac{\tilde{w}_n}{v} - \frac{1}{\gamma_n}, 0\},
\end{equation}
where $v$ is determined by $\sum_{n=1}^{KN_S}\lambda_n = \sum_{n=1}^{KN_S} \max\{\frac{\tilde{w}_n}{v} - \frac{1}{\gamma_n}, 0\} = KN_S$. This solution is a revised version of the traditional water-filling power allocation, which takes the weights of all MSs into account, termed as proportional water-filling. An insight from (\ref{eq:proportional_wf}) can be interpreted as variable water-levels: one MS with a greater weight $\tilde{w}_n = w_k$ has a higher water-level $\frac{\tilde{w}_n}{v}$, where more power can be allocated, and vice verse.

\subsection{Performance Analysis in ULA Single-Path Channels}\label{sec:perform_analysis}
Due to the discretization of receive vectors in analog combiners and the baseband block diagonalization, analyzing the sum spectral performance of the hybrid BD scheme is indeed non-trivial. Nevertheless, it is tractable to present the performance analysis of a special case with ULA single-path channels and large numbers of transmit and receive antennas ($N_{BS}, N_{MS} \to \infty$). Note that, in the mmWave channels, both the BS and MSs need to employ large antenna arrays to harvest adequate receive power from the signals passing through a few propagation paths \cite{alkh2014limited}. To conduct the performance analysis, we impose the following assumption that each MS only schedules one data stream through the only one RF chain, which is $N_S = M_{MS} = 1$, while the BS is equipped with $M_{BS} = K$ RF chains. Herein, the single-path channel for the $k$-th MS in (\ref{eq:cl_ch_model}) can be rewritten as
\begin{equation}\label{eq:single_ch_model}
    \mathbf{H}_k = \sqrt{N_{BS}N_{MS}}\alpha^k \mathbf{a}_{MS}^k(\theta^k) \mathbf{a}_{BS}^k(\phi^k)^H,
\end{equation}
where $\alpha_k$ is the result of the large scale path fading $\sqrt{\beta_k}$ multiplied the complex gain of this unique path, while $\mathbf{a}_{MS}^k(\theta^k)$ and $\mathbf{a}_{BS}^k(\phi^k)$ are the corresponding receive and transmit array response vectors. Besides, the analog combiner has only one column, denoted as $\mathbf{W}_k = \mathbf{w}_k$. With a large number of receive antennas $N_{MS}$, the candidates of DFT bases in problem (\ref{eq:wk_pro_dft_cons}) will have an infinite resolution. Under this circumstance, we have
\begin{equation}\label{eq:wk_dft_equip}
\begin{aligned}
    &\max_{\mathbf{W}_k}~\sum_{m=1}^{M_{MS}}||({\mathbf{w}_k^{(m)}})^H\mathbf{H}_k||_1^2 = \max_{\mathbf{w}_k}~ ||\mathbf{w}_k^H\mathbf{H}_k||_1^2 \\
    & = \max_{\mathbf{w}_k}~||\sqrt{N_{BS}N_{MS}}\alpha^k [\mathbf{w}_k^H\mathbf{a}_{MS}^k(\theta^k)] \mathbf{a}_{BS}^k(\phi^k)^H||_1^2 \\
    & = \max_{\mathbf{w}_k}~\{\sqrt{N_{BS}N_{MS}}\alpha^k [\mathbf{w}_k^H\mathbf{a}_{MS}^k(\theta^k)] \cdot ||\mathbf{a}_{BS}^k(\phi^k)^H||_1\}^2 \\
    & = \max_{\mathbf{w}_k}~\{N_{BS}\sqrt{N_{MS}}\alpha^k [\mathbf{w}_k^H\mathbf{a}_{MS}^k(\theta^k)]||_1\}^2.
\end{aligned}
\end{equation}
Therefore, the analog combiner for the $k$-th MS should be approximately $\mathbf{w}_k \approx \mathbf{a}_{MS}^k(\theta^k)$, selected from the DFT bases of infinite resolution when $N_{MS} \to \infty$. 

Furthermore, the entries in the baseband equivalent channel $\mathbf{H}_{eq}$ can be determined through applying EGT. As we define an operator $\mathsf{g}(\cdot)$ that imposes the element amplitudes of the input vector as unit, the $(k,j)$-th entry of $\mathbf{H}_{eq}$ is given by
\begin{equation}\label{eq:heq_kj}
\begin{aligned}
    \mathbf{H}_{eq}^{(k,j)} &= \sqrt{\frac{1}{N_{BS}}}\left[ \mathbf{w}_k^H\mathbf{H}_k \cdot \mathsf{g}((\mathbf{w}_j^H\mathbf{H}_j)^H) \right], \\
    & = \sqrt{N_{BS}N_{MS}}\alpha^k \mathbf{a}_{BS}^k(\phi^k)^H \cdot \\
    &~~~~~~~~~~~~~~\sqrt{\frac{1}{N_{BS}}}\mathsf{g}(\sqrt{N_{BS}N_{MS}}\alpha^j \mathbf{a}_{BS}^j(\phi^j)) \\
    & = \sqrt{N_{BS}N_{MS}}\alpha^k \mathbf{a}_{BS}^k(\phi^k)^H \cdot \mathbf{a}_{BS}^j(\phi^j).
\end{aligned}
\end{equation}
With the form of $N_{BS}$-element ULA antenna setting, it is intuitive that $\mathbf{a}_{BS}^k(\phi^k)^H \mathbf{a}_{BS}^k(\phi^k) = 1$, while
\begin{equation}\label{eq:abs_product_kj}
\begin{aligned}
    &{[\mathbf{a}_{BS}^k(\phi^k)^H \mathbf{a}_{BS}^j(\phi^j)]}_{k \neq j} \\
    &~~~= \frac{1}{N_{BS}}\sum_{n=0, k \neq j}^{N_{BS}-1}e^{j\frac{2\pi}{\lambda}nd(\sin{\phi^j} - \sin{\phi^k})} \\
    &~~~= \frac{1}{N_{BS}} \frac{1 - e^{\frac{2\pi}{\lambda}d(\sin{\phi^j} - \sin{\phi^k})N_{BS}}}{1 - e^{\frac{2\pi}{\lambda}d(\sin{\phi^j} - \sin{\phi^k})}}.
\end{aligned}
\end{equation}
Without the loss of generality, we regard that $\sin{\phi^j} - \sin{\phi^k} \neq 0$ as long as $k \neq j$. Then we safely draw a conclusion that 
\begin{equation}\label{eq:offdiag_infinitesmal}
\begin{aligned}
    \lim_{N_{BS} \to \infty}\frac{\mathbf{H}_{eq}^{(k, j)}}{\mathbf{H}_{eq}^{(k, k)}} &= \lim_{N_{BS} \to \infty}\mathbf{a}_{BS}^k(\phi^k)^H \mathbf{a}_{BS}^j(\phi^j) \\
    & = \lim_{N_{BS} \to \infty}\frac{1}{N_{BS}} \frac{1 - e^{\frac{2\pi}{\lambda}d(\sin{\phi^j} - \sin{\phi^k})N_{BS}}}{1 - e^{\frac{2\pi}{\lambda}d(\sin{\phi^j} - \sin{\phi^k})}} \\
    & = 0,
\end{aligned}
\end{equation}
where $k \neq j$. Therefore, the baseband equivalent channel can be approximated as a diagonal matrix after analog precoding and combining, given by $\mathbf{\Lambda}_{eq} = \sqrt{N_{BS}N_{MS}}\mathrm{diag} \cdot \{\alpha^1, \alpha^2, \cdots, \alpha^K \}$, due to the fact that the off-diagonal entries are infinitesimal compared with the diagonal entries as $N_{BS}, N_{MS} \to \infty$. There is no need to do block diagonalization except the water-filling power allocation at the baseband to achieve the optimal sum spectral efficiency as 
\begin{equation}\label{eq:approx_cap_single_ula}
    R \approx \log_2 \left( \left| \mathbf{I}_{K} + \frac{P\mathbf{\Lambda}{\mathbf{\Lambda}_{eq}}^2}{\sigma^2K}\right| \right),
\end{equation}
where $\mathbf{\Lambda}$ is a diagonal matrix that performs water-filling power allocation. Eq. (\ref{eq:approx_cap_single_ula}) is an approximate sum spectral efficiency under the settings of ULA single-path channels and large number of transmit and receive antennas, and we will present this analytical result in the simulations.

\section{Simulation Results}\label{sec:simulation}
In this section, we evaluate the spectral efficiency achieved by the Hy-BD scheme as well as its performance robustness in the massive MU-MIMO channels. 

\subsection{Spectral Efficiency Evaluation}
In the simulations of this section, we illustrate the spectral efficiency achieved by the Hy-BD scheme in the massive MU-MIMO systems by comparing it with the traditional high-dimensional baseband BD scheme in large i.i.d Rayleigh fading and mmWave multiuser channels and also with the previously proposed spatially sparse precoding/combining scheme \cite{heath2014sparse} in mmWave channels. The range of the signal-to-noise ratio $\mathrm{SNR} = \frac{P}{\sigma^2}$ is from -40 dB to 0 dB in all processing solutions. And the large-scale fading path loss factor $\beta_k, k=1,\cdots, K,$ of all MSs are uniformly distributed in $[0.5, 1.5]$. All MSs have equal unit weights in the simulations.

\begin{figure}[htbp]
    \centering
    \includegraphics[width=0.45\textwidth]{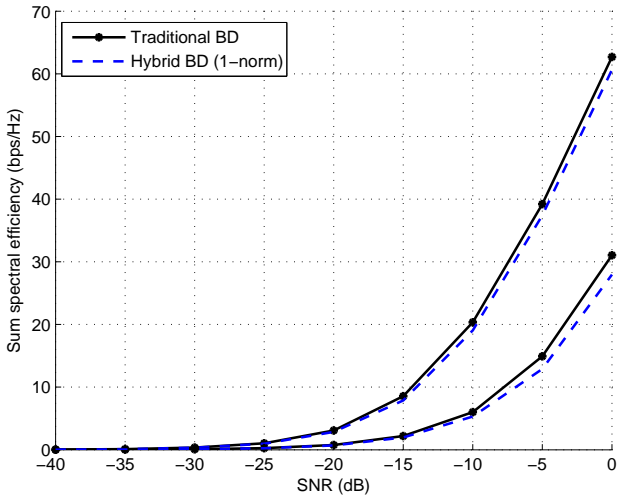}
    \caption{Sum spectral efficiency achieved by different processing schemes in an 8-user MU-MIMO system in i.i.d. Rayleigh fading channels where $N_S=2, M_{MS} = 2, M_{BS} = 16$.}
    \label{fig:cap_rayleigh}
\end{figure}

Fig. \ref{fig:cap_rayleigh} illustrates the sum spectral efficiency achieved by the traditional BD scheme and our proposed Hy-BD scheme in the large i.i.d. Rayleigh fading channel. The BS with $M_{BS} = 16$ RF chains is employed to schedule $K=8$ MSs, each of which processes $N_S = 2$ data streams with $M_{MS} = 2$ RF chains. Furthermore, the BS and MSs are equipped with 256 (16) and 64 (4) antennas respectively. In both $256 \times 16$ and $64 \times 4$ antenna settings, the sum spectral efficiency of the Hy-BD scheme consistently approaches the performance achieved by the traditional BD scheme, however, with lower implementation and computational complexity. Notably, the results of the $64 \times 4$ antenna setting indicate that the Hy-BD scheme is still effective in a small scale antenna system.

\begin{figure}[htbp]
\centering
\begin{subfigure}{0.45\textwidth}
    \centering
    \includegraphics[width=\linewidth]{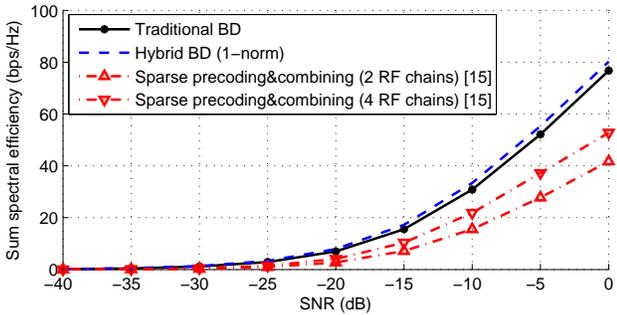}
    \caption{Performance with ULA}
    \label{fig:cap_mmwave_ula_upa1}
\end{subfigure}
\begin{subfigure}{0.45\textwidth}
    \centering
    \vspace{0.5em}
    \includegraphics[width=\linewidth]{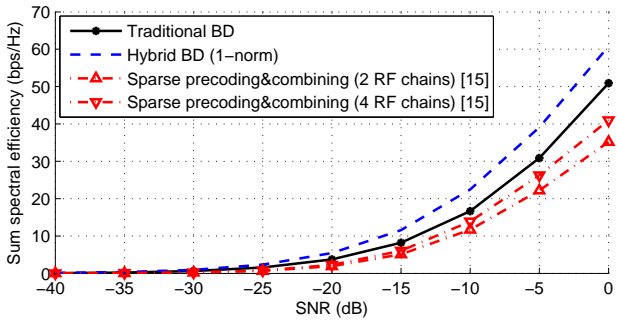}
    \caption{Performance with UPA}
    \label{fig:cap_mmwave_ula_upa2}
\end{subfigure}
    \caption{Sum spectral efficiency achieved by different processing schemes in an $256 \times 16$ 8-user MU-MIMO system in mmWave channels where $N_S=2, M_{MS} = 2(4), M_{BS} = 16(32)$.}
    \label{fig:cap_mmwave_ula_upa}
\end{figure}

In the mmWave MU-MIMO channels, the traditional full-complexity BD and Hy-BD schemes perform in a similar fashion as in the Rayleigh fading channels. Based on the limited number of paths scattered in the mmWave channels, the spatially sparse precoding/combining scheme in \cite{heath2014sparse} can be extended to the hybrid processing in MU-MIMO systems through decomposing the solution to the traditional BD scheme (the precoder $\mathbf{M}_S$ and the MMSE combiners in \cite{spencer2004bd}) via orthogonal matching pursuit where the BS and MSs choose the array response vectors of the corresponding AoDs and AoAs as the basis vectors respectively. Fig. \ref{fig:cap_mmwave_ula_upa} shows the sum spectral efficiency of the above processing schemes with ULA and UPA employed respectively.\footnote{Under the UPA setup, it is necessary to introduce extra elevation angle for each propagation paths. In the simulations, we use the same settings for both elevation and azimuth angles.}
We set the mmWave propagation channel with $N_c = 8$ and $N_p = 10$. The range of the mean azimuth angles of AoDs at the BS $|\theta_\mathrm{max}^k - \theta_\mathrm{min}^k|$ is $120^\circ$ while the MSs are assumed to be omni-directional due to the relatively smaller antenna array elements. The angle spreads $\sigma_\theta^k$'s and $\sigma_\phi^k$'s are all equal to $7.5^\circ$ (the settings of azimuth angles are also applied to elevation angles in UPA setup). Moreover, the BS is set to have $N_{BS} = 256$ antennas and $M_{BS} = 16$ RF chains, while $K=8$ MSs, with $N_{MS} = 16$ antennas and $M_{MS} = 2$ RF chains, all dealing with $N_S = 2$ data streams. 
In this scenario, the proposed Hy-BD scheme even achieves slightly higher spectral efficiency than the traditional BD scheme, while the performances of the Hy-BD scheme and spatially sparse coding scheme are upgraded when the system applies UPA instead of ULA. Note that the traditional BD scheme is a sub-optimal solution for the processing of MU-MIMO systems, and it is possible that the Hy-BD outperforms the traditional BD in some situations. As for the spatially sparse precoding/combining scheme, it lags behind the traditional BD and Hy-BD schemes because the columns of the traditional BD precoding and combining matrices do not directly come from the linear combination of the array response vectors of AoDs/AoAs, the basic forming units of the RF matrices in the spatially sparse coding scheme \cite{heath2014sparse}. This is very different from the P2P scenario that the spatially sparse scheme is designed for, where the columns of the SVD based precoder and combiner can be effectively approached by the linear combinations of the array response vectors according to the observation 3) in \cite{heath2014sparse}. Even though the number of RF chains is enlarged to $M_{MS} = 4$ and $M_{BS} = 32$, the performance of the spatially sparse precoding/combining scheme is still inferior to the full-complexity BD and Hy-BD schemes.

\begin{figure}[htbp]
\centering
\begin{subfigure}{0.45\textwidth}
    \includegraphics[width=\linewidth]{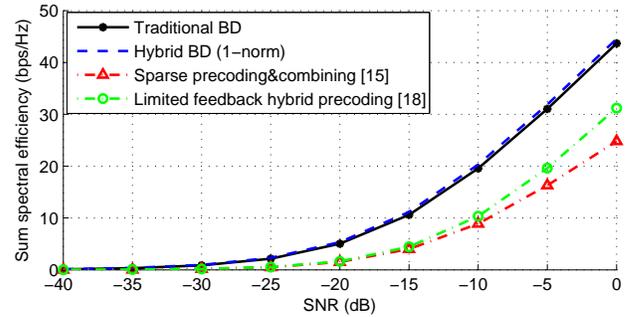}
    \caption{Performance with ULA}
    \label{fig:cap_mmwave_lf_ula_upa1}
\end{subfigure}
\begin{subfigure}{0.45\textwidth}
    \centering
    \vspace{0.5em}
    \includegraphics[width=\linewidth]{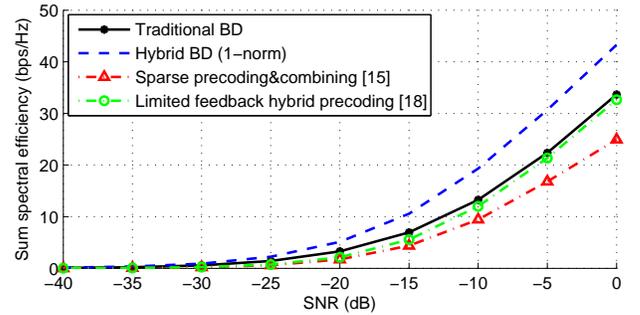}
    \caption{Performance with UPA}
    \label{fig:cap_mmwave_lf_ula_upa2}
\end{subfigure}
    \caption{Sum spectral efficiency achieved by different processing schemes in an $256 \times 16$ 8-user MU-MIMO system in mmWave channels where $N_S=1, M_{MS} = 1, M_{BS} = 8$.}
    \label{fig:cap_mmwave_lf_ula_upa}
\end{figure}

\begin{figure}[htbp]
    \centering
    \includegraphics[width=0.45\textwidth]{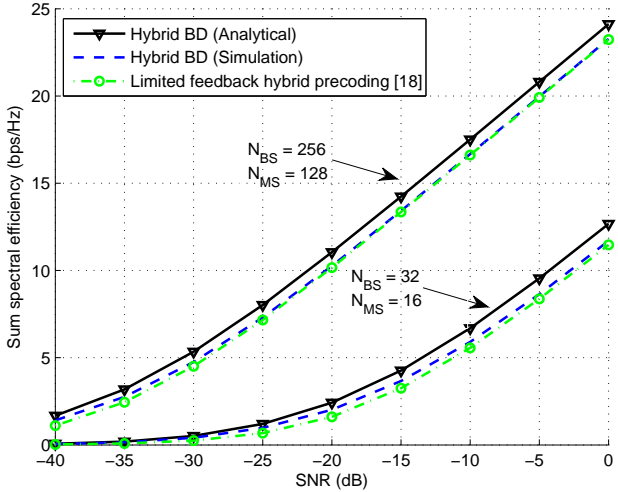}
    \caption{Sum spectral efficiency achieved by different processing schemes in ULA single-path channels where $M_{MS} = N_S = 1, M_{BS} = K = 2$.}
    \label{fig:cap_analytical}
\end{figure}

Considering the critical situation that only one data stream is supported by each MS with only RF chain employed (total 8 MSs), we are able to further compare our results with the limited feedback hybrid precoding scheme proposed in \cite{alkh2014limited} in Fig. \ref{fig:cap_mmwave_lf_ula_upa}. It shows that the proposed Hy-BD scheme still outperforms other baselines. Although the limited feedback hybrid precoding scheme is capable of tracking the strongest path in the mmWave channels, it fails to harvest the large array gain when the mmWave channel for each MS is not extremely sparse since only an RF chain pair is available for each MS to track one propagation path in the RF domain (we generate 80 paths for each MS's mmWave channel in the simulations of Fig. \ref{fig:cap_mmwave_lf_ula_upa}). However, with the Hy-BD scheme, the EGT enabled by the RF precoder can directly aggregate the channel gains so that the spectral efficiency performance can be guaranteed. Furthermore, the approximate sum spectral efficiency of hybrid BD scheme in ULA single-path channels, analyzed in Section \ref{sec:perform_analysis} is illustrated in Fig. \ref{fig:cap_analytical}, where $M_{MS} = N_S = 1$ and $M_{BS} = K = 2$. It shows that the hybrid BD performs closely to its analytical approximate version, with about 1 bps/Hz degradation, which is caused by limited numbers of transmit and receive antennas as well as the DFT restriction. In this circumstance, the hybrid BD scheme and limited feedback method obtain similar performance since they are both capable of tracking the channel's unique path.

\subsection{Robustness Evaluation}
In addition to simply demonstrating the spectral efficiency of the Hy-BD scheme under different SNRs, we further examine its performance robustness by changing the multiplexing settings (e.g., the number of data streams supported by each user and the number of users) and introducing the channel estimation error.

\begin{figure}[htbp]
    \centering
    \includegraphics[width=0.45\textwidth]{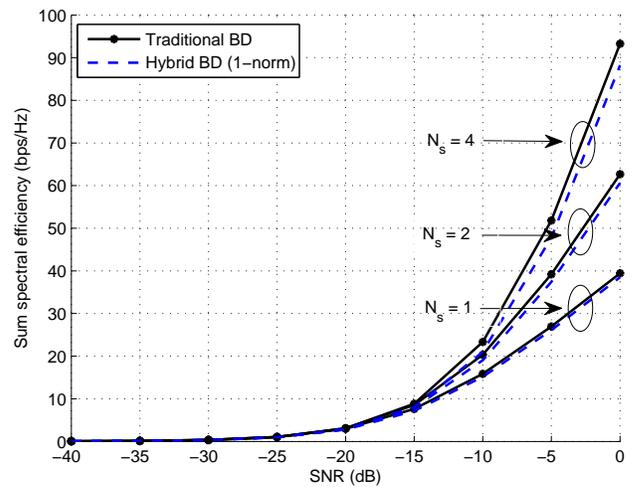}
    \caption{Sum spectral efficiency achieved by different processing schemes in an 8-user MU-MIMO system in i.i.d. Rayleigh fading channels where $N_S = M_{MS} =1,2,4$ and $M_{BS} = 8M_{MS}$.}
    \label{fig:cap_rayleigh_k8_ns124}
\end{figure}

\begin{figure}[htbp]
    \centering
    \includegraphics[width=0.45\textwidth]{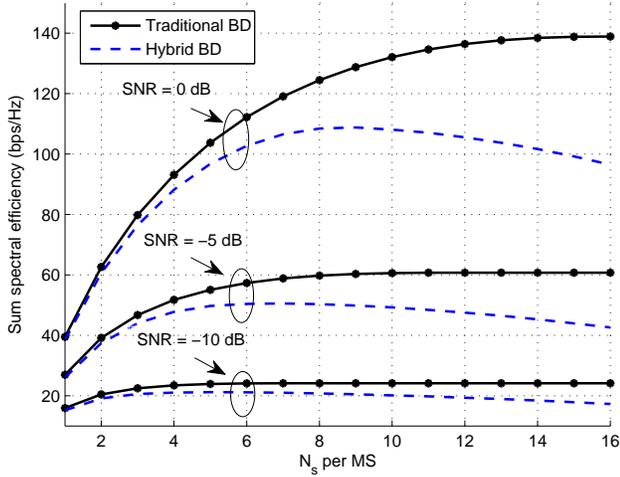}
    \caption{Sum spectral efficiency achieved by different processing schemes in a $256 \times 16$ 8-user MU-MIMO system in i.i.d. Rayleigh fading channels where $N_S$ increases from 1 to 16 and SNR = $-10, -5, 0$ dB.}
    \label{fig:cap_rayleigh_k8_ns1_16}
\end{figure}

\begin{figure}[htbp]
    \centering
    \includegraphics[width=0.45\textwidth]{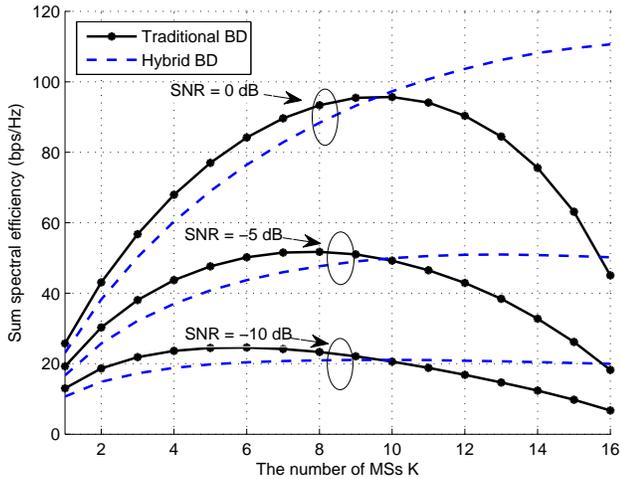}
    \caption{Sum spectral efficiency achieved by different processing schemes in a $256 \times 16$ MU-MIMO system in i.i.d. Rayleigh fading channels where $K$ increases from 1 to 16, SNR = $-10, -5, 0$ dB and $N_S = 4$.}
    \label{fig:cap_rayleigh_k1_16_ns4}
\end{figure}

For the practical implementation of an MU-MIMO system, the total number of supported data streams is a very important criterion to evaluate the system performance, which depends on the number of supported MSs $K$ and the number of data stream supported by each MS $N_S$, namely, space-division multiple access and spatial multiplexing. In Figs. \ref{fig:cap_rayleigh_k8_ns124}-\ref{fig:cap_rayleigh_k1_16_ns4}, the sum spectral efficiency achieved by the traditional BD scheme and the Hy-BD scheme is checked in a $256 \times 16$ 8-user MU-MIMO system in i.i.d Rayleigh fading channels under different SNRs, where each MS only employs $M_{MS} = N_S$ RF chains ($K*N_S$ RF chains at the BS) in the Hy-BD scheme.

In Fig. \ref{fig:cap_rayleigh_k8_ns124}, the number of data streams per MS is set as $N_S = 1, 2, 4$ and the SNR ranges from -40 dB to 0 dB. The gap between the sum spectral efficiency of the traditional BD scheme and the Hy-BD scheme remains minute compared to the absolute sum spectral efficiency. However, the Hy-BD scheme only needs the same number of RF chains as the supported data streams at both BS and MSs (up to $M_{BS} = 32$ and $M_{MS} = 4$), much smaller than that of the traditional BD scheme ($M_{BS} = 256$ and $M_{MS} = 16$). 
Fig. \ref{fig:cap_rayleigh_k8_ns1_16} shows the sum spectral efficiency of both schemes when $N_S$ increases from 1 to 16 and the SNR is set as $-10, -5$ and $0$ dB, which indicates that it is suitable to employ the Hy-BD scheme when the total number of data streams in the MU-MIMO system is not too large, so that the Hy-BD can reach the peak spectral efficiency. As we can see, the sum spectral efficiency achieved by the traditional BD scheme will be continuously augmented in such a $256 \times 16$ 8-user MU-MIMO system when the number of transmitted data streams increases, since more equivalent parallel channels (characterized by the diagonal elements in $\mathbf{\Sigma}$ in \cite{spencer2004bd}) can be utilized to transmit the data streams and the effect of inter-user interference is not dominant in this case. However, in the Hy-BD scheme, the spectral efficiency performance is somewhat compromised once a large quantity of data streams are transmitted. This is because the pursuit of the large array gain slightly introduces the inter-stream interference in the RF domain, which will degrade the system spectral efficiency after the baseband BD processing. On the other hand, with an increasing SNR, the suitable numbers of the supported data streams $N_S$, corresponding to the peak spectral efficiency, for the traditional BD scheme and Hy-BD scheme are also enhanced. For instance, when SNR = $0$ dB, the traditional BD scheme supports up to $K*N_S = 8*16 = 128$ data streams which is the maximum number of the supported data streams by a $256 \times 16$ 8-user MU-MIMO system with full RF chains. However, the Hy-BD scheme can support about $K*N_S = 8*8 = 64$ data streams with only 64 and 8 RF chains at the BS and MS respectively.

With the same system configuration as that of Figs. \ref{fig:cap_rayleigh_k8_ns124} and \ref{fig:cap_rayleigh_k8_ns1_16}, and the number of data streams per MS set as $N_S = 4$, the number of MSs $K$ increases from 1 to 16 in Fig. \ref{fig:cap_rayleigh_k1_16_ns4}. In this case, the traditional BD scheme with full RF chain configuration reaches a peak spectral efficiency at a certain $K$. This is because when $K$ grows beyond an optimal value, inter-user interference substantially becomes more severe and the sum spectral efficiency is gradually degraded. As for the Hy-BD scheme with the limited RF chain configuration, the sum spectral efficiency keeps improving when $K$ increases from 1 to 16 (the maximum number of supported data stream is still up to $K*N_S = 4*16 = 64$). By comparing the results of Figs. \ref{fig:cap_rayleigh_k8_ns1_16} and \ref{fig:cap_rayleigh_k1_16_ns4}, the Hy-BD scheme can be safely recommended for implementation in  systems with a large number of MSs, however, each of which deals with a small number of data streams, since it is less vulnerable to the inter-user interference than the traditional BD scheme in this case. As for the case in Fig. \ref{fig:cap_rayleigh_k8_ns1_16} where there are fewer MSs and more data streams per MS, the traditional BD scheme achieves superior performance at the cost of high complexity because it can better process the inter-stream interference than the Hy-BD scheme.

\begin{figure}[htbp]
    \centering
    \includegraphics[width=0.45\textwidth]{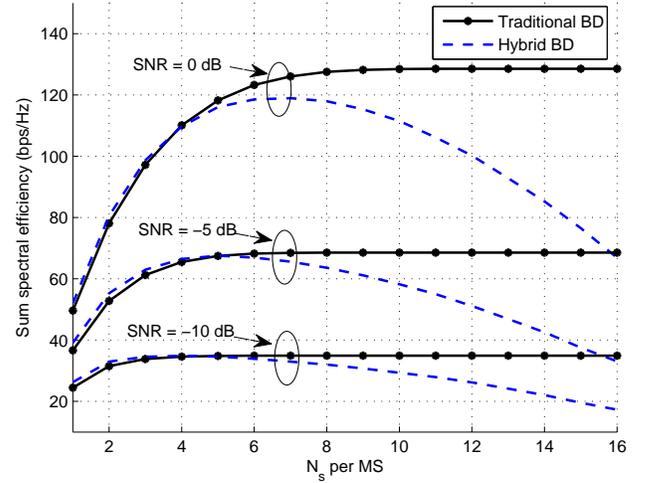}
    \caption{Sum spectral efficiency achieved by different processing schemes in a $256 \times 16$ 8-user MU-MIMO system in mmWave channels where $N_S$ increases from 1 to 16 and SNR = $-10, -5, 0$ dB.}
    \label{fig:cap_mmwave_k8_ns1_16}
\end{figure}

\begin{figure}[htbp]
    \centering
    \includegraphics[width=0.45\textwidth]{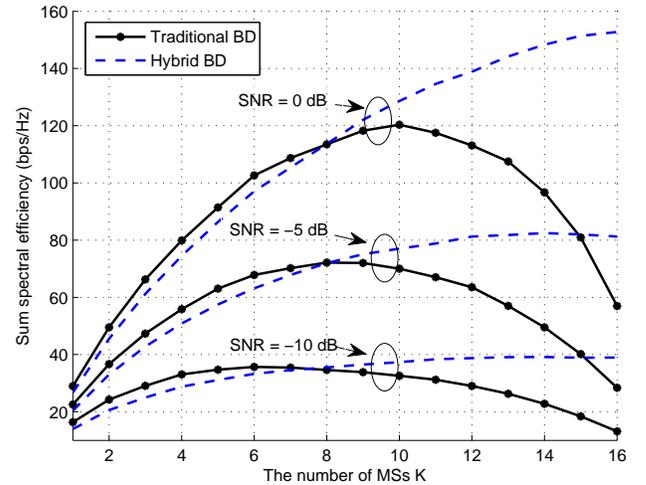}
    \caption{Sum spectral efficiency achieved by different processing schemes in a $256 \times 16$ MU-MIMO system in mmWave channels where $K$ increases from 1 to 16, SNR = $-10, -5, 0$ dB and $N_S = 4$.}
    \label{fig:cap_mmwave_k1_16_ns4}
\end{figure}

Furthermore, we examine the sum spectral efficiency of both schemes with an increasing $N_S$ or $K$ under different SNRs ($-10, -5$ and $0$ dB) in the mmWave MU-MIMO channels whose propagation characteristics are given in Fig. \ref{fig:cap_mmwave_ula_upa}'s settings. The BS and MS configurations are the same as those of Figs. \ref{fig:cap_rayleigh_k8_ns1_16} and \ref{fig:cap_rayleigh_k1_16_ns4}. Here, Fig. \ref{fig:cap_mmwave_k8_ns1_16} illustrates the sum spectral efficiency of both schemes when $N_S$ increases from 1 to 16 with $K = 8$, while Fig. \ref{fig:cap_mmwave_k1_16_ns4} gives the result for the number of MSs $K$ increasing from 1 to 16 with $N_S = 4$. As can be seen, the general trends of the sum spectral efficiency of the traditional BD scheme and the Hy-BD scheme in mmWave channels are consistent with those in Rayleigh fading channels, except that the Hy-BD scheme can perform slightly better in mmWave channels compared with the results in Rayleigh fading channels. It is probably due to the fact that the DFT bases selection (conforming to the forms of AoAs/AoDs array responses of the limited number of paths in mmwave channels) in the Hy-BD scheme essentially captures the dominant paths of the mmWave channels. 

\section{Conclusion}
In this paper, a low-complexity hybrid block diagonalization processing scheme has been proposed for the downlink communication of a massive multiuser MIMO system with the limited number of RF chains. We harvest the large array gain through the phase-only RF precoding and combining and then the BD technique is performed at the equivalent baseband channel. It has been demonstrated that the Hy-BD scheme, with a lower implementation and computational complexity, achieves a capacity performance approaching that of the traditional high-dimensional baseband BD processing. Such a low-complexity, low cost Hy-BD scheme can be a promising option for the practical implementation of a massive MU-MIMO system.

%





\begin{thebibliography}{99}
\bibitem{andrews2014what5g}
J.\ G.\ Andrews, S.\ Buzzi, C.\ Wan, S.\ V.\ Hanly, A.\ Lozano, A.\ C.\ K.\ Soong, and J.\ C.\ Zhang, ``{What Will 5G Be},'' \emph{IEEE Journal on Selected Areas in Commun.}, vol.\ 32, pp.\ 1065--1082, June 2014.

\bibitem{rusek2013scalemimo}
F.\ Rusek, D.\ Persson, B.\ K.\ Lau, E.\ G.\ Larsson, T.\ L.\ Marzetta, O.\ Edfors, and F.\ Tufvesson, ``{Scaling up MIMO: Opportunities and challenges with very large arrays,}'' \emph{IEEE Sig. Process. Mag.}, vol.\ 30, pp.\ 40–60, Jan.\ 2013.

\bibitem{larsson2014massive}
E.\ G.\ Larsson, O.\ Edfors, F.\ Tufvesson, and T.\ L.\ Marzetta, ``{Massive MIMO for next generation wireless systems},'' \emph{IEEE Commun. Mag.}, pp.\ 186–195, vol.\ 52, no.\ 2, Feb.\ 2014.

\bibitem{marzetta2010noncoop_mimo}
T.\ L.\ Marzetta, ``{Noncooperative cellular wireless with unlimited numbers of base station antennas},'' \emph{IEEE Trans. on Wireless Commun.}, vol.\ 9,\ pp.\ 3590--3600, Nov.\ 2010.

\bibitem{erez2005dpc}
U.\ Erez, S.\ Shamai and R.\ Zamir, ``{Capacity and lattice strategies for canceling known interference},'' \emph{IEEE Trans. on Info. Theory}, vol.\ 51, pp.\ 3820--3833, Nov.\ 2005.

\bibitem{spencer2004bd}
Q.\ H.\ Spencer, A.\ L.\ Swindlehurst, and M.\ Haardt, ``{Zero-forcing methods for downlink spatial multiplexing in multiuser MIMO channels},'' \emph{IEEE Trans. on Sig. Process.}, vol.\ 52, pp.\ 461--471, Feb. 2004.

\bibitem{hanif2014beam}
M.\ F.\ Hanif, Le-Nam Tran, A.\ Tolli and M.\ Juntti, ``{Computationally efficient robust beamforming for SINR balancing in multicell downlink with applications to large antenna array systems},'' \emph{IEEE Trans. on Commun.}, vol.\ 62, pp.\ 1908-1920, June 2014.

\bibitem{hanif2014conic}
Le-Nam Tran, M.\ F.\ Hanif and M.\ Juntti, ``{A conic quadratic programming approach to physical layer multicasting for large-scale antenna arrays},'' \emph{Signal Process. Letters, IEEE}, vol.\ 21, pp.\ 114-117, Jan. 2014.

\bibitem{peng2010las}
P. Li and R.\ D. Murch, ``{Multiple output selection-LAS algorithm in large MIMO systems},'' \emph{Commun. Letters, IEEE}, vol.\ 14, pp.\ 399-401, May 2010.


\bibitem{love2003equalgain}
D.\ J.\ Love and R.\ W.\ Heath, ``{Equal gain transmission in multiple-input multiple-output wireless systems},'' \emph{IEEE Trans. on Commun.}, vol.\ 51, pp.\ 1102–1110, July 2003.

\bibitem{zhang2005antselect}
X.\ Zhang, A.\ F.\ Molisch, and S.\ Y.\ Kung, ``{Variable-phase-shift-based RF-baseband codesign for MIMO antenna selection},'' \emph{IEEE Trans. on Sig. Process.}, vol.\ 53, pp.\ 4091–4103, Nov. 2005.

\bibitem{liang2014hybrid}
L.\ Liang, W. Xu and X.\ Dong, ``{Low-Complexity Hybrid Precoding in Massive Multiuser MIMO Systems},'' \emph{IEEE Wireless Commun. Letters}, Oct. 2014.

\bibitem{liu2014rfonly}
A.\ Liu, and V.\ Lau, ``{Phase only RF precoding for massive MIMO systems with limited RF chains},'' \emph{IEEE Trans. on Sig. Process.}, vol.\ 62, pp.\ 4505--4515, Sept. 2014.


\bibitem{heath2012beamsteer}
O.\ E.\ Ayach, R.\ W.\ Heath, S.\ Abu-Surra, S.\ Rajagopal and Z. Pi, ``{The capacity optimality of beam steering in large millimeter wave MIMO systems},'' in \emph{Proc. IEEE 13th Intl. Workshop on Sig. Process. Advances in Wireless Commun. (SPAWC)}, pp.\ 100--104, June 2012.

\bibitem{heath2014sparse}
O.\ E.\ Ayach, S.\ Rajagopal, S.\ Abu-Surra, Z.\ Pi and R.\ W.\ Heath, ``{Spatially sparse precoding in millimeter wave MIMO systems},'' \emph{IEEE Trans. on Wireless Commun.}, vol.\ 13, pp.\ 1499–1513, Mar. 2014.

\bibitem{sayeed2013beamspace}
A.\ Sayeed and J.\ Brady, ``{Beamspace MIMO for high-dimensional multiuser communication at millimeter-wave frequencies},'' in \emph{Proc. IEEE Global Commun. Conf. (GLOBECOM)}, pp.\ 3679--3684, Dec. 2013.

\bibitem{striling2014hybrid}
R.\ A.\ Stirling-Gallacher and Md.\ S. Rahman, ``{Linear MU-MIMO pre-coding algorithms for a millimeter wave communication system using hybrid beam-forming},'' in \emph{Proc. IEEE Intl. Conf. on Commun. (ICC)}, pp.\ 5449--5454, June 2014.

\bibitem{alkh2014limited}
A.\ Alkhateeb, G.\ Leus and R.\ W.\ Heath\ Jr, ``{Limited feedback hybrid precoding for multi-user millimeter wave systems},'' available in \emph{arXiv:1409.5162}, 2014.

\bibitem{weighted2008christensen}
S.\ S.\ Christensen, R.\ Agarwal, E.\ Carvalho and J.\ M.\ Cioffi, ``{Weighted Sum-Rate Maximization using Weighted MMSE for MIMO-BC Beamforming Design},'', \emph{IEEE Trans. on Wireless Commun.}, vol.\ 7, pp. 4792-4799, Dec. 2008.

\bibitem{dinesh2012csadaption}
D.\ Ramasamy, S.\ Venkateswaran and U.\ Madhow, ``{Compressive adaptation of large steerable arrays},'' in \emph{Proc. Info. Theory and App. Workshop (ITA)}, pp.\ 234--239, Feb. 2012.

\bibitem{boyd2004opt}
S.\ Boyd and L.\ Vandenberghe, \emph{Convex Optimization}, Cambridge University Press, 2004.


\end{thebibliography}
\end{document}